\documentclass[article,twocolumn]{revtex4-2}
\usepackage{natbib}
\usepackage{graphicx}
\usepackage{ dsfont }
\usepackage{color}
\definecolor{darkblue}{rgb}{0.0,0.0,0.5}
\usepackage[colorlinks,citecolor=darkblue,linkcolor=darkblue,allcolors=darkblue]{hyperref}
\usepackage{amsmath}
\definecolor{darkgreen}{rgb}{0.2,0.7,0}

\DeclareMathOperator{\sinc}{sinc}
\newcommand{\vac}{|{\rm vac}\rangle}
\newcommand{\ket}[1]{\left| #1 \right\rangle}
\newcommand{\bra}[1]{\left\langle #1 \right|}
\newcommand{\braket}[2]{\left\langle #1 | #2 \right\rangle}
\newcommand{\expect}[1]{\left\langle#1\right\rangle}
\newcommand{\proj}[1]{| #1\rangle\!\langle #1 |}
\newcommand{\Tr}{\mathrm{Tr}}
\newcommand{\fff}{\mathrm{post}}
\newcommand{\dd}{\mathrm{d}}

\newcommand{\NN}{\mathds{N}}
\newcommand{\ZZ}{\mathds{Z}}
\newcommand{\CC}{\mathds{C}}
\newcommand{\RR}{\mathds{R}}
\newcommand{\LL}{{\cal L}}
\newcommand{\eea}{\end{eqnarray}}
\newcommand{\bea}{\begin{eqnarray}}
\newcommand{\ee}{\end{equation}}
\newcommand{\be}{\begin{equation}}

\begin{document}
\title{All Hilbert spaces are the same: consequences for generalized coordinates and momenta}

\author{S.~J.~van Enk}
\email{svanenk@uoregon.edu}
\author{Daniel A.~Steck} % orcid: 0000-0002-9120-7650
\email{dsteck@uoregon.edu}
\affiliation{Department of Physics and
Oregon Center for Optical, Molecular \& Quantum Sciences\\
University of Oregon, Eugene, OR 97403}
\begin{abstract}
Making use of the simple fact that all separable complex Hilbert spaces of given dimension are isomorphic, we show that  there are just six basic ways to define generalized coordinate operators in Quantum Mechanics. In each case a canonically conjugate generalized momentum operator can be defined, but it may not be self-adjoint. Even in those cases we show there is always either a self-adjoint extension of the operator or a Neumark extension of the Hilbert space that produces a self-adjoint momentum operator. In one of the six cases both extensions work, thus leading to seven basic pairs of coordinate and momentum operators.
We also show why there are more ways of defining basic coordinate and momentum measurements. A special role is reserved for measurements that simultaneously but imperfectly measure both.
\end{abstract}
\maketitle
\section{Introduction}

Every infinite-dimensional separable complex Hilbert space $H$ is isomorphic to $l^2(\NN)$, the Hilbert space consisting of all sequences of complex numbers $x:=(x_n)$ with $\sum_n |x_n|^2<\infty$, with inner product
$(x,y)=\sum_n x^*_n y_n$. The label $n$ here (and in the next sentence) ranges over the natural numbers, $n\in\NN$.
For each separable Hilbert space $H$ we can define a countable orthonormal basis set $\{\ket{n}\}$ and then associate $(x_n)\in l^2$ with the vector $\sum_n x_n \ket{n}\in H$.  The isomorphism between $l^2$ and $H$ is thus completely determined by the specification of the basis $\{\ket{n}\}$  for $H$ \footnote{All this uses physicists' notation, namely, Dirac's bra-ket notation. Since there are well-documented dangers in Dirac's formalism \cite{gieres2000mathematical}, in the following we will, when appropriate, indicate such dangers. See Ref.~\cite{hall2013quantum} and Ref.~\cite{talagrand2022quantum} for excellent pedagogical expositions of the rigorous mathematics behind quantum mechanics. Both books also provide a guide to translating physics literature, when possible (which is not an empty proviso), to rigorous results.}. An example of such a countable basis for $L^2(\RR,\dd x)$ consists of the eigenfunctions of the simple 1-dimensional harmonic oscillator (for {\em any} positive mass $m$ and positive frequency $\omega$).

As our starting point we assume we have constructed a separable Hilbert space to describe a given quantum system (see the Appendix for more information about such constructions in, e.g., quantum field theory).  
Since such a Hilbert space is typically infinite-dimensional and thus isomorphic to $l^2(\NN)$, it may be useful to state what determines the {\em physical} differences between different physical systems so described. 
 The answer consists of two parts: First, every (closed) system has a specific self-adjoint operator associated with it that describes the time evolution of the system. This operator, of course, is the Hamiltonian. Second, the Hamiltonian is defined in terms of more elementary operators. From classical physics we borrow pairs of generalized coordinate and momentum variables that are canonically conjugate to each other and that become self-adjoint operators in the quantum description. 
 The Hamiltonian may further contain other operators that do not have classical counterparts, such as the spin operator for an electron, to mention one example.  
 
 Here our focus will be entirely on the pairs of generalized coordinate and momentum operators. Given that there is only one type of Hilbert space within which we define such operators, it may not be surprising that our choice is limited. In Section II we show there are six basic types of coordinate operators. In five cases this leaves one type of self-adjoint momentum operator. In one exceptional case two types are possible. Thus there are seven basic pairs of coordinate and momentum operators.

In Section III we show that measurements of position and momentum for a given quantum system are not fixed by just defining their corresponding operators. Rather, the coupling of the system to a measurement device determines what exactly is measured. Considering this coupling will quite generically lead to measurements of both position and momentum, each with a finite nonzero accuracy, constrained by standard uncertainty relations.

We include an Appendix with 9 sections in total. The main aim is to collect for convenience relevant and more technical results that are spread out over a large mathematical and physical literature. We also include some new illustrative examples there, 
which illustrate the physical consequences of our work.
For example, we consider the tricky case of the moments of momentum for a
constant wave function, for a particle on a finite interval:
are they zero, or do they diverge?
We also consider the case of calculating $\langle[X^4,P^4]\rangle$
for a particle in a box on $[-L/2,L/2]$: should the answer be
zero or not?

 \section{Coordinate and momentum operators}
  Our discussions of momentum operators in this Section will make use of the concepts of {\em self-adjoint extensions} and {\em Neumark extensions}.
 Pedagogical introductions to both concepts aimed at physicists can be found in  Refs.~\cite{gieres2000mathematical,bonneau2001self,garbaczewski2004impenetrable,brandt1999positive,juric2022}.  A very simple example of a Neumark 
(sometimes transliterated as Naimark) extension is given in the second section of the Appendix. The third section defines the difference between ``hermitian'' and ``self-adjoint'' operators. Its physical import is illustrated with simple examples in the fourth section of the Appendix.

 We assume we have a model that is meant to (approximately) describe some physical system within 
 some restricted domain. We assume  
 we have constructed an appropriate Lagrangian $\LL$, defined in terms of generalized coordinates $q_j$ and velocities $\dot{q}_j$. We then eliminate velocities in favor of momenta $p_j$ by moving to a Hamiltonian description, in which $p_j=\partial \LL/\partial \dot{q}_j$ and the Hamiltonian is
 ${\cal H}(\{q_j,p_j\})=\sum_j p_j\dot{q}_j-\LL$. 
 
 Sometimes our coordinates and momenta thus obtained are clearly distinct types of physical quantities. For example, we may have invariance under translations of the coordinate, which then implies that its associated momentum is conserved (but not the other way around).
 Sometimes it makes physical sense to {\em a priori} restrict the coordinate to be positive, but we would not restrict momentum in that way. Such (intuitive) asymmetries between the coordinates and momenta typically manifest themselves in the form of the Hamiltonian. This explains why our constructions in the following will not be symmetric between coordinates and momenta. 
 
 For a simple harmonic oscillator, however, we do have symmetry between $q$ and $p$. In the case of the free electromagnetic (EM) field, modeled as a set of independent harmonic oscillators (called field ``modes''), both $p$ and $q$ for a given mode are field quadrature variables which do not differ in character from each other (as long as the field is not interacting). 

 We restrict ourselves to analyzing 1-dimensional systems on a line and their basic Hilbert spaces. This is justified by noting that Hilbert spaces for more complicated systems can be obtained either by taking tensor products of such basic Hilbert spaces, or direct sums, or combinations thereof. Thus a standard $N$-dimensional system may be described by an $N$-fold tensor product, while topologically more complicated 1-dimensional systems, say a tree graph with $N$ edges, can be described as an $N$-fold direct sum. The same holds for a union of $N$ disjoint intervals.

 We thus focus on a single generalized coordinate and momentum pair $(q=:x,p)$ (the fifth section of the Appendix discusses $\RR^3$).
We denote the corresponding quantum operators by $X$ and $P$.
 We need to make two choices for the coordinate $x$. 
 The first choice concerns whether $x$ is continuous or discrete. The second choice is about the interval over which $x$ is defined: the interval could be bounded on both sides, bounded on one side, or not bounded in either direction. Thus there are $6=2\times 3$ basic choices to be made for fixing the coordinate operator. 
 We emphasize that these 6 choices are by no means meant to all make physical sense for every physical system. Rather, if we vary over all possible physical models, then these are the 6 possible basic choices. 
 Once we have made our choice for the coordinate operator $X$, we then consider in each case what possibilities exist for defining a self-adjoint (canonically conjugate) momentum operator $P$.
 As is well known, boundary conditions are typically used to adjust the domain of the momentum operator so that it may become both hermitian and self-adjoint. Boundary conditions typically do not  affect the choice of Hilbert space.

 Since the standard position operator on $L^2(\RR,\dd x)$ has a continuous spectrum, we start with assuming $x$ is a continuous coordinate.
 
 \subsection{Continuous $x$}
 Given a continuous variable $x$, there are three further choices possible, as to whether the interval over which $x$ is defined is (1) infinite, (2) semi-infinite (bounded on one end, unbounded on the other) or (3) finite.
 We use the following three subsets $\Omega_k$ of $\RR$ for $k=1,2,3$ to exemplify these three (topologically different) cases:
 \bea
 \Omega_1&=&\RR,\nonumber\\
 \Omega_2&=& [0,\infty),\nonumber\\
 \Omega_3&=& [0,L],
 \eea
  for some fixed real value of $L>0$. 
  The corresponding Hilbert spaces are $H_k=L^2(\Omega_k,\dd x)$, consisting of the square-integrable complex functions on $\Omega_k$, with inner product $(f,g)=\int_{\Omega_k} f^*(x) g(x) \dd x$.
  In each case we define the coordinate operator as
    \be
 X_k=\int_{\Omega_k}  x\proj{x} \dd x.
 \ee 
 Here $\dd\mu_x:=\proj{x}\dd x$ is an operator-valued measure.
 That is,
 $\int_{\Omega_k} \dd\mu_x =\openone_{H_k}$, with the right-hand side denoting the identity operator on the Hilbert space $H_k$. It is, moreover, a projection-operator valued measure, which means the following. For any measurable subset $E$ of $\Omega_k$, define $\mu_E:=\int_E \proj{x}\dd x$. Then, for any two such subsets $E,E'$ that are disjoint we have
 $\mu_E \mu_{E'}=0$. (That this is so, follows, in physicists' notation, from $\braket{x'}{x}=\delta(x'-x)$.) 
It is thanks to this last property that all three operators $X_k$ defined above are self-adjoint. That is, $X_k^\dagger=X_k$ and their domains coincide (see \cite{gieres2000mathematical,bonneau2001self,hall2013quantum}).
 
 Whereas $X_3$ is a bounded operator and thus defined on the whole Hilbert space $H_3$, $X_1$ and $X_2$ are unbounded and thus only defined on a dense subspace of $H_{1,2}$ \cite{hall2013quantum}. 
 For example, the so-called Schwartz space consisting of all rapidly decreasing smooth ``test functions'' is a dense subspace of $L^2(\RR,\dd x)$ \cite{becnel2015schwartz}. On the finite interval, the set of polynomial functions forms a dense subspace of $H_3$ \footnote{In an uncountably infinite space, incidentally, the 
  Schwartz space $\mathcal{S}$ can be used to make Hilbert space
  rigorous, in the form of a rigged Hilbert space.
  First define the so-called Gel'fand triple,
  $\mathcal{S}\subset L^2(\RR)\subset\mathcal{S}'$, where $\mathcal{S}'$
  is the space of tempered distributions \cite{Gelfand64}.
  Then, for example, the position eigenstate in the position
  representation $\langle x|x'\rangle = \delta(x-x')$ is in $\mathcal{S}'$, in the case of continuous $x$.
  Inner products involving elements of $\mathcal{S}'$ are then defined
  with elements of $\mathcal{S}$ only.
  For example, if $H=L^2(\RR)$, then the Hilbert space inner product is
  $\braket{f}{g}=\int \dd x \, f^*(x)g(x)$ with both $f$ and $g$ elements of $H$. If, however, we restrict $g$ to be an element of the smaller space $\mathcal{S}$ then $f$ may be chosen from the larger space $\mathcal{S'}$ while still keeping the integral well-defined. 
  (This way we never encounter problematic or undefined integrals like
  $\int \dd x\,\delta^2(x)$.) 
  After defining this, other parts of Dirac notation are made
  more rigorous, too.  For example, for the operator-valued measure
  $\dd\mu_x=\proj{x}\dd x$ we use in the text, for $\psi,\phi\in\mathcal{S}$, the inner product involving this operator is $\bra{\psi}\dd\mu_x\ket{\phi}=\psi^*(x)\phi(x)\dd x$.}
 
 It is tempting to consider Cases 2 and 3 as special cases of Case 1, in which one or two infinite barriers are used to model one or two high but finite barriers that prevent our quantum system from escaping subregion $\Omega_{2,3}$, respectively. The operators $X_2$ and $X_3$ may then simply be considered as restrictions of the operator $X_1$ to smaller domains.
 For the associated momentum operators $P_k$, however, the situation is not so simple \cite{garbaczewski2004impenetrable}.

 We wish to define a momentum operator canonically conjugate to the operator $X_k$ for $k=1,2,3$.
 For that purpose we use the Fourier transform (rather than 
 directly trying to define an operator like $-i \dd/\dd x$; we set $\hbar=1$ here).
 In physicists' notation we first define for any $p\in\RR$
 \be\label{p}
 \ket{p}:=\frac{1}{\sqrt{2\pi}} \int_{\Omega_k} \exp(ipx) \ket{x} \dd x, 
 \ee
 where we integrate over the appropriate range $\Omega_k$ for $x$.
 We note that neither the symbol $\ket{x}$ nor the symbol $\ket{p}$ stand for vectors in $H_k$. (See Ref.~\cite{talagrand2022quantum} for the general strategy to convert an expression like (\ref{p}) to a rigorously defined expression. Our end results are well-defined.)
 
Importantly,  independently of the domain $\Omega_k$  we have
 \be\label{POM}
 \int_\RR  \proj{p} \dd p=\int_{\Omega_k} \proj{x} \dd x =\openone_{H_k}
  \ee
  for all three cases $k=1,2,3$.
That is, $\dd\mu_p:=\proj{p}\dd p$ defines an operator-valued measure in each case.

In all three cases $k=1,2,3$ we may then define a candidate momentum operator $P_k$ as
 \be
 P_k=\int_\RR  p\proj{p} \dd p.
 \ee
Whether this operator is self-adjoint depends on the range of integration over $x$. We consider the three cases $k=1,2,3$ separately.
\subsubsection{Case 1}
When the domain of $x$ is the entire real axis,  $\dd \mu_p=\proj{p} \dd p$ is a projection-valued measure. 
That this is so, follows, in physicists' notation, from $\braket{p'}{p}=\delta(p'-p)$. 
This implies that $P_1$ is self-adjoint. In this case, we do simply have $P_1=-i \dd/\dd x$.
The commutator $[X_1,P_1]=i\openone_{H_1}$ is thus the standard one.
Arguably the more fundamental relation is the exponentiated (Weyl) form of the commutator, in terms of unitary displacement operators,
\be\label{W}
%\exp(itX_1)\exp(isP_1)=\exp(-ist)\exp(isP_1)\exp(it X_1)
e^{isP_1}e^{it X_1}=e^{itX_1}e^{isP_1}e^{its},
\ee
for all $s,t\in\RR$. This relation is stronger than the standard commutation relation
and implies it.
The Stone-von Neumann theorem states \cite{hall2013quantum,summers2001stone} that any pair of operators $X_1,P_1$ that satisfy (\ref{W}) for all  $s,t\in\RR$ and that are weakly continuous in $s,t$ are unitarily equivalent to the standard position and momentum operators on $L^2(\RR,\dd x)$. 
Each assumption going into the Stone-von Neumann theorem can be revoked. See Chapter 3 of Ref.~\cite{Ruetsche2011} for representations that may result in those cases.
Later we will see that in finite-dimensional Hilbert spaces the Weyl relation survives in discretized form (see Eq.~(\ref{dWeyl})).

 \subsubsection{Case 2}\label{section:case2}
For the semi-infinite axis,  $P_2$ is not self-adjoint, as the measure  $\dd\mu_p=\proj{p} \dd p$ is not projection-operator valued. 
That is, there are disjoint subsets $E$ and $E'$ of $\RR$ such that 
\be
\left(\int_E \proj{p} \dd p \right) \left (\int_{E'} \proj{p} \dd p\right) \neq 0. 
\ee
This follows, again in physicists' notation, from the presence of the second term here: 
\be\braket{p'}{p}=\frac{1}{2}\delta(p'-p)+\frac{i}{2\pi(p-p')}.
\ee
On the other hand, $\mu_p$ is a positive-operator valued measure (POVM, see (\ref{POM})). By Neumark's theorem, the POVM $\mu_p$ on $H_2$ can be extended to a projection-operator valued measure on some larger Hilbert space $H'$ \cite{helstrom1969quantum,brandt1999positive,wang2023generalized}.
  As such, it represents a generalized measurement on the Hilbert space $H_2$, which can be interpreted as arising from a restriction to $H_2$ of a standard projective measurement on the larger Hilbert space $H'$. 
  It also means that the operator $P_2$ can be viewed as a restriction of a self-adjoint momentum operator $P'_2$ on $H'$.

 There are two ways of choosing the larger Hilbert space $H'$. First, we could obviously choose $H'=H_1=L^2(\RR, \dd x)$. Second, we could choose  $H'=L^2((0,\infty), \dd x)\otimes \CC^2$ \cite{al2021canonical}.
 In both cases we have ``doubled'' the total Hilbert space: in the first case by extending the semi-infinite real axis to the whole real axis, in the second case by using an extra (``ancilla'' in the jargon of quantum information theory) qubit described by a complex vector space $\CC^2$ with standard inner product.
 
 The main physical difference between these two Neumark extensions shows up when we consider measurements. In the first case the quantum system's state after the measurement is no longer confined to its original Hilbert space $H_2$---for example, during the measurement the potential barrier at $x=0$ was lowered or even removed. In the second case the system's state is still inside the original Hilbert space---but the state of the ancilla qubit may have changed. (And so a pure initial state of our system may have been changed into a mixed state.) 
 We return to measurements in Section~\ref{MM}.
 Furthermore, we give a simple toy example of Neumark's extension in the Appendix to show its end result can be interpreted in (at least) two different ways if the Hilbert space is (at least) doubled.

Note we included the measure $\dd x$ explicitly in the definition of our Hilbert space $H_2=L^2(\Omega_2,\dd x)$. This is important because the semi-infinite axis can, of course, be mapped onto the whole real axis by mapping $x>0$ to $y=\ln(x)$. This would give a different Hilbert space $H_2'=L^2(\Omega_2,\dd x/x)$.
On that different Hilbert space we can define a canonically conjugate operator by defining
$P'=-ix \dd/\dd x$. But this would simply be equivalent to $P_1$ as defined on $H_1$.

 We end this subsection by noting that there exists no self-adjoint extension of the momentum operator on the semi-infinite axis. That is, we cannot construct boundary conditions at $x=0$ such that the domain of $P_2$ becomes equal to the domain of the adjoint operator $P_2^\dagger$. 
 This follows from a theorem by von Neumann in terms of ``deficiency indices'' as discussed in Refs.~\cite{bonneau2001self,gieres2000mathematical}.
 That there is no such self-adjoint extension for the semi-infinite axis is most easily understood intuitively by noting there cannot be a unitary spatial translation operator, which would map the positive real axis reversibly to itself.

 \subsubsection{Case 3}
For the finite interval case, just as in Case 2,  the measure  $\dd\mu_p=\proj{p} \dd p$ is not projection-operator valued, although it is a positive-operator valued measure. The operator $P_3$ is, therefore, not self-adjoint. In this case, both types of extensions work. That is, we can define a self-adjoint extension of $P_3$ on $H_3$ and we can define a Neumark extension by enlarging our Hilbert space to $H_1$. One may even combine the two methods and define a self-adjoint extension on a larger (e.g., doubled) Hilbert space \cite{al2021alternative}. 

If we extend the Hilbert space $H_3$ to $L^2(\RR, \dd x)$, momentum is continuous, just as position is.
The other choice for defining the momentum operator yields an operator with discrete eigenvalues, labeled by integers $m\in \ZZ$. Specifically, we may define
\be\label{pmv}
p_m=\frac{2\pi m}{L}+\frac{\theta_0}{L},
\ee
where $\theta_0$ can be chosen within the interval $[0,2\pi)$. 
We define then
\be\label{pm}
\ket{p_m(\theta_0)}:=\frac{1}{\sqrt{L}}\int_{\Omega_3}  \exp(ip_m x)\ket{x} \dd x.
\ee
We have, as is easily verified,
\be
\sum_{m\in \ZZ}  \proj{p_m}=\int_{\Omega_3} \proj{x} \dd x=\openone_{H_3}.
\ee
The alternative choice for a momentum operator is then
\be\label{P3theta}
\tilde{P}_3(\theta_0):=\sum_{m\in \ZZ}  p_m \proj{p_m}.
\ee
It is self-adjoint because the vectors $\{\ket{p_m}\}$ form an orthonormal basis set. 
The result obtained is, in fact, just the standard family of self-adjoint extensions of the operator $-i \dd/\dd x$ parametrized by the $U(1)$ parameter $\theta_0$ \cite{gieres2000mathematical,bonneau2001self,garbaczewski2004impenetrable}. 
Henceforth we suppress the argument $\theta_0$ in definitions (\ref{pm}) and (\ref{P3theta}).

 That there are self-adjoint extensions for the finite interval is most easily understood intuitively by noting one can construct a unitary spatial translation operator by making it ``wrap around'' the interval $[0,L]$.
 This wrap-around translation operator is generated by the self-adjoint momentum operator $\tilde{P}_3$ (by Stone's theorem \cite{hall2013quantum,talagrand2022quantum}). 
 Even though the commutator of $X_3$ and $\tilde{P}_3$ is the standard one,
 the exponentiated (Weyl) form differs from (\ref{W}), precisely because translations in  position have to wrap around the interval. In fact, there is an additional phase factor \cite{hall2013quantum,talagrand2022quantum}
 \bea
  e^{is \tilde{P}_3}e^{it X_3}
 =e^{it X_3}e^{is \tilde{P}_3}e^{it(s-m_{x,s}L)},
 \eea
when acting on a wave function $\psi(x)$ with $x\in [0,L]$.
Here $m_{x,s}$ is the unique integer such that
\be
  0\leq x+s-m_{x,s}L <L.
\ee
The extra phase factor  
$\exp(-it m_{x,s}L)$ 
does not always equal 1.
This way, the Stone-von Neumann theorem 
says nothing about
coordinate and momentum operators on a finite interval.

Note incidentally that the uncertainty principle here is more complicated than the standard one for continuous position and momentum
\cite{judge1963,judge1964,bouten1965,evett1965}.

In conclusion, for the finite interval there are two possible types of momentum operators, one continuous, the other discrete, one obtained from a Neumark extension, the other as a self-adjoint extension.
These two extensions apply to physically (and topologically) different situations. The Neumark extension corresponds to the infinite square well seen as a limit of a finite square well. The self-adjoint extension with $\theta_0=0$ corresponds to a particle moving on a circle (or a cylinder), since in this case $x=0$ and $x=L$ are identified. The case $\theta_0\neq 0$ corresponds to a charged particle moving around a circle (or cylinder) that encloses a nonzero magnetic flux. 
Comparing the two extensions, 
a Neumark extension involves an expansion of the Hilbert space,
while self-adjoint extensions do not.

 \subsection{Discrete $x$}
 If we choose our coordinate $x$ to be discrete,  our second choice concerns the labeling of these discrete values. We could choose to label them by integers $m \in\ZZ$ (Case 4), by natural numbers $n\in\NN$ (Case 5), or by a finite set of $D$ natural numbers (Case 6). The corresponding Hilbert spaces are different here than in the preceding subsections. For the finite case it is simply $H_6=\CC^D$, the second discrete case gives us $H_5=l^2(\NN)$, and the Hilbert space for first case may be denoted as $H_4=l^2(\ZZ)$ with the only difference with $l^2(\NN)$ being that sequences of complex numbers are indexed by an integer in $\ZZ$.  
 \subsubsection{Case 4}
 Case 4, with $x$ unbounded in two directions, leads to the most straightforward definition of momentum (just like it did for the continuous case).
 
 We may choose the eigenvalues of the operator $X_4$ to be $x_m=m L_0 + \lambda_0$, with $L_0$ a constant with units of length and $0\leq \lambda_0 < L_0$. We may, for example, choose $L_0$ to equal the Planck length. In this way, we can be sure no experiment has ever been performed that might contradict conclusions drawn from just this choice, $L_0$ being far smaller than any spatial resolution  ever achieved.  
 (This description takes the operator $X_4$ literally as indicating position in units of length, but it is easy to amend the description to other cases, where the coordinate $x$ is, for example, an angle or  a field quadrature. One just needs the value of $L_0$ to be much smaller than the accuracy with which the corresponding physical quantity has ever been measured.)
 Our coordinate operator is  
 \be
 X_4=\sum_{m\in\ZZ} x_m\proj{x_m}.
 \ee
 We choose our candidate momentum operator by again using the Fourier transform. We first follow the physicists' way by defining ``delta-function normalized'' kets
 \be\label{pd}
 \ket{p}=\sqrt{L_0}\sum_{m\in\ZZ} \exp(i px_m)\ket{x_m},
 \ee
 and we hasten to add that  $\ket{p}$ is not an element of our Hilbert space $H_4$ (unlike $\ket{x_m}$).
 We may restrict $p$ to an interval of length $2\pi/L_0$, without loss of generality. Symmetry suggests choosing $p\in\Omega_p:=[-\pi/L_0,\pi/L_0]$. 
The measure $\dd\mu_p$ is projection-operator valued so that
 \be
 P_4=\int_{\Omega_p} p \proj{p}\, \dd p
 \ee
 is self-adjoint. We see from this description that, in fact, our Hilbert space can also be taken as $H_4=L^2(\Omega_p, \dd p)$.
 We have
 \be
 \int_{\Omega_p}  \proj{p}\, \dd p=\sum_{m\in \ZZ} \proj{x_m}=\openone_{H_4}.
 \ee
 The result here is very similar to the self-adjoint extension obtained for Case 3 with $\theta_0=2\pi \lambda_0/L_0$, but with the roles of $x$ and $p$ interchanged.

  \subsubsection{Case 5}
 Now we choose $x_n=n L_0$ with $n\in \NN$.  This case naturally describes a quantum system moving on the semi-infinite real axis. Our fifth position operator is then
  \be
 X_5=\sum_{n\in\NN} x_n\proj{x_n}.
 \ee
Just as in Case 4 we start in physicists' notation with  
 \be
\ket{p}=\sum_{n\in\NN} \exp(i px_n)\ket{x_n}.
\ee
While we do have
 \be
\int_{\Omega_p}  \proj{p}\, \dd p=\sum_{n\in \NN} \proj{x_n}=\openone_{H_5},
\ee
the operator-valued measure $\dd\mu_p =\proj{p} \dd p$ is not a projection-operator valued measure because $\braket{p}{p'}\neq \delta(p-p')$ (half of the Fourier sum is missing).
The operator 
\be
 P_5=\int_{\Omega_p} p \proj{p}\, \dd p
 \ee
 is, therefore, not self-adjoint.
 $\mu_p$ does describe a POVM and thus a generalized measurement. By Neumark's theorem 
 we can extend (``double'') the Hilbert space either by reverting to Case 4 or by using an ancilla qubit as in Case 2. In fact, Case 5 is similar (for obvious reasons) to Case 2, the semi-infinite axis.
 \subsubsection{Case 6: finite-dimensional Hilbert space}

The finite-dimensional case is straightforward since the coordinate and momentum operators will both be bounded. The coordinate operator may be used to describe different sorts of discrete physical coordinates.
We could use a discretized angle $\phi$ as generalized coordinate and hence angular momentum as our canonically conjugate ``momentum'' operator. We could also use a finite number of number states counting energy units $\hbar\omega$ in an electromagnetic field mode, and then use the phase of the field as conjugate variable. We could also use a quantum wall clock with integers indicating the whole hours; the conjugate variable would then be the angular momentum of the hour hand. We now mimic the procedure followed in Cases 1--5 to construct operators $X_6$ and $P_6$.

For any of the specific physical systems just mentioned, suppose our coordinate basis states $\ket{n}$ are labeled by $D$ consecutive integers $n=n_0, n_0+1,\ldots n_0+D-1$ (some or all of which could be negative). Thus, $H_6=\CC^D$.
We define then a continuum of (``Fourier-transformed'' but not unit-length) vectors $\ket{\phi}\in H_6$ for real $\phi$ as
\be\label{phik}
\ket{\phi}=
\sum_{n} \exp(in\phi)\ket{n}.
\ee
We may restrict the domain of $\phi$ to some interval of length $2\pi$, which we do not have to specify at this point yet.
Then we have
\be
\frac{1}{2\pi}
\int
\proj{\phi}\, \dd \phi=\sum_n \proj{n}=\openone_H.
\ee
Two different vectors $\ket{\phi}$ and $\ket{\phi'}$ are not in general orthogonal. We have
\be
\braket{\phi'}{\phi}=\sum_n \exp(in(\phi-\phi')).
\ee
This equals zero if and only if $\phi$ and $\phi'$ differ by an integral multiple of $2\pi/D$. Thus, we can define an alternative orthogonal basis for $H_6$ by using $D$ vectors
$\ket{\phi_n}$ with $\phi_n=n 2\pi/D+\phi_0$ with $n=n_0,n_0+1,\ldots n_0+D-1$ and some arbitrary $\phi_0$.
We may then define our generalized coordinate and momentum operators as
\bea
P_6&=&\sum_n \phi_n \frac{\proj{\phi_n}}{D},\nonumber\\
X_6&=&\sum_n n \proj{n}. 
\eea
In a finite-dimensional Hilbert space, $X_6$ and $P_6$ lose their interpretation as generators, because infinitesimal displacements in these coordinates are not allowed.  Instead the finite-displacement operators become ${\cal X}:=\exp(-2\pi iP_6/D)$ and ${\cal Z}:=\exp(2\pi iX_6/D)$ (which for $D=2$ are the Pauli operators $\sigma_x$ and $\sigma_z$, respectively, hence the notation. 
The operators ${\cal X}$ and ${\cal Z}$
satisfy the (discrete) Weyl form of the commutation relation
\cite{vourdas2004quantum}
\be \label{dWeyl}
{\cal X}^n{\cal Z}^m=\exp(-2\pi mn i/D) {\cal Z}^m{\cal X}^n,
\ee
for any integers $m,n$.
This relation gives the cleanest connection to the standard position and momentum operators defined on $L^2(\RR,\dd x)$; compare Eq.~(\ref{W}).

 \subsubsection{Cases 4 and 5 revisited}
 We assumed in the preceding subsections that the labeling
 preserves the spatial order of the coordinates $x_n$. That is, 
 whenever $x_n<x_k<x_{n'}$ we have $n<k<n'$.
 Although this is very natural it is not necessary, and there is one reason for not following this prescription in Case 5 where $n\in \NN$. Namely, since we needed a Neumark extension to define a self-adjoint momentum operator, whereas in Case 4 we did not, we may simply find a one-to-one mapping between $\NN$ and $\ZZ$ and thus map Case 5 onto Case 4. One of the simpler ways to define this mapping is to map even natural numbers to nonnegative integers and odd natural numbers to negative integers. That is, we use this map:
 \bea
 n\mapsto m&=&n/2 \,\,{\rm for}\,\, n\in\NN \,\,{\rm even},\nonumber\\
 n\mapsto m&=&-(n+1)/2  \,\,{\rm for}\,\, n\in\NN \,\,{\rm odd}.
 \eea
 This mapping provides a different point of view on the procedure followed in Ref.~\cite{al2021canonical} for defining a self-adjoint momentum operator on the semi-infinite axis by first distinguishing even and odd points on a discrete lattice and then taking the continuum limit. 
 \subsubsection{Discrete position on a finite interval revisited}
 In Case 6 we defined a finite number of discrete coordinates in a finite interval. We may, however, define infinitely many of them. In that case we need to let $|x_{n+1}-x_n|$ go to zero for $|n|\rightarrow\infty$. This case is not really physically different from the Cases 4 or 5, however. Namely, we just interpret $x$ as merely a coordinate and declare the physical distance between positions $x_n$ and $x_{n'}$ to depend only on $|n-n'|$, e.g., by setting $||x_n-x_{n'}||=|n-n'|L_0$. In this way we just have equipped the interval with a metric, such that it becomes physically equivalent to Case 4 if $n\in \ZZ$ or to Case 5 if $n\in \NN$. (For completeness, we note that the choice of rational coordinates is technically possible, too, but it would (drastically) not preserve spatial order.)

 \section{Measurements of position and momentum}\label{MM}
Now we turn to the subject of measurements.
We first discuss the physical requirements of measurements.
In considering POVM measurements of both position and momentum,
we focus on the ``finest'' measurements, discussing the meaning
of this in terms of the purity of the POVM element.
We then discuss how these coherent-state measurements
fit into our Cases 1-6 developed above.

 \subsection{Preliminary remarks}
For Cases 1--5 either the coordinate operator or the momentum operator (in Cases 1 and 2, both) has no eigenvalues and no eigenstates in Hilbert space. 
That is, they \textit{have} eigenvalues and eigenstates, 
but they are generalized eigenvalues and eigenstates,
and they lie outside Hilbert space \cite{gieres2000mathematical}.
Thus, the usual link suggested between measurements of observables and eigenstates and eigenvalues of corresponding operators does not work for the coordinate and momentum operators for Cases 1--5.

 Our coordinate and momentum operators do form the kinematical basis of quantum descriptions, but for dynamics, including measurements, we need a Hamiltonian \footnote{In some cases the construction of a physical separable Hilbert space as an appropriate subspace of an initial non-separable Hilbert space for infinitely many degrees of freedom, alluded to in the Appendix on QFT, depends explicitly on the dynamics, namely, when the choice between unitarily inequivalent representations is informed by the dynamics.}.
Measurements of coordinates and momenta are determined by
 the possible interaction Hamiltonians that describe how our quantum system is coupled to a measurement device. We may, therefore, expect there to be a larger variety of possible measurements than there are operators for just the system.
 
  That the coordinate operator does not uniquely determine measurements is illustrated by the fact that even if $x$ is continuous, we may still assume our coordinate measurements to be discrete and coarse-grained with some finite resolution $L_0$. (The converse is true, too; see below.)
   Such a  discrete measurement could, e.g., be defined as projecting onto the orthonormal set of states
 \be\label{condis}
 \ket{m}=\frac{1}{\sqrt{L_0}}\int_{x_m}^{x_{m+1}}  \ket{x} \dd x,
 \ee
 where $x_m=mL_0+\lambda_0$, which span a subspace of $H_1$, naturally isomorphic to $H_4$ or $H_5$.
  Note, however, how this definition of $\ket{m}$ differs from Cases 4 and 5. 
   Several variations on the definition (\ref{condis}) would be possible as well.
 For example, if we allow variations in both $m$ and $p$ in the following definition
  \be
 \ket{p,m}=\frac{1}{\sqrt{L_0}}\int_{x_m}^{x_{m+1}} \exp(ipx) \ket{x} \dd x,
 \ee 
 then projectors onto this overcomplete set of nonorthogonal states would constitute a POVM (extendible by Neumark's theorem to a standard measurement on a larger Hilbert space) that describes simultaneous measurements of $x$ with resolution $L_0$ and of $p$ with resolution on the order of $1/L_0$. 

This sort of possibility---a simultaneous measurement of two non-commuting observables---is much more than a curiosity. According to Ref.~\cite{she1966simultaneous} an actual measurement of position always also measures, albeit with very low resolution, momentum, and an actual measurement of momentum always also measures, with very low resolution, position.
The argument for this is simple: a measurement of momentum on a quantum system takes place inside some measurement device inside some lab, and so the position of the system, both before and after the measurement, is limited to some finite range.
Similarly, an actual measurement of position of a quantum system leaves the kinetic energy of the system bounded and so momentum squared and thus momentum itself are bounded, too.

This train of thought can be extended to yield a few reasonable physical requirements on descriptions of an actual measurement.
Suppose a measurement (of whatever physical quantity) projects onto some post-measurement state (which may be a mixed state $\rho$ or a pure state $\ket{\psi}$) of our system. A reasonable requirement is that $\langle P^2\rangle_{\fff}$, the expectation value of $P^2$ in the post-measurement state, be bounded. In fact, it seems reasonable to require that $\langle P^{2m}\rangle_{\fff}$ be bounded for any fixed positive integer $m$. (Note that repeated measurements of $P$ on identically prepared quantum systems suffice to estimate higher-order moments $\expect{P^n}$.)
Similarly, if we choose our coordinate $x$ such that $x=0$ somewhere inside our measurement setup, then $\langle X^{2m}\rangle_{\fff}$ ought to be bounded as well.

Joint measurements of incompatible observables are necessarily described by a POVM, not by a projection-valued measure. In fact, since the state in which our system is left after a measurement plays an important role, we ought to consider Kraus operators \cite{kraus1983states} $K_\alpha$ for each possible outcome labeled $\alpha$ (which may be continuous or discrete or multi-index). This operator determines the map from any pre-measurement state $\rho$ to the post-measurement state $\rho'$ for outcome $\alpha$ as
\be
\rho\longmapsto \rho'= K_\alpha \rho K_\alpha^\dagger /\sqrt{P_\alpha},
\ee
where $P_\alpha=\Tr(K_\alpha^\dagger K_\alpha \rho)$ is the probability of obtaining measurement result $\alpha$. Note that the relation between Kraus operator and POVM element $\Pi_\alpha$ is $\Pi_\alpha=K_\alpha^\dagger K_\alpha$.
(In the Appendix we discuss Kraus operators as they apply to a continuous measurement of both $X_1$ and $P_1$ and how that description connects to the next topic, the coherent-state measurement.)

We impose one more constraint on our descriptions of basic measurements by focusing on the ``finest'' possible measurements.
That is, we do not include POVMs that are ``coarser'' than other POVMs.
For example, the Kraus operator for the Gaussian position measurement
\begin{equation}
  K_{x_0,w} = \left(\frac{1}{2\pi w}\right)^{1/4}
  \int_{\RR}e^{-(x-x_0)^2/4w}\ket{x}\!\!\bra{x}\, \dd x
\end{equation}
is a coarser measurement than the coherent-state measurement,
\begin{equation}
  K_{\alpha} = \frac{1}{\sqrt{\pi}}\ket{\alpha}\!\!\bra{\alpha},
  \label{coherentstatemeasurement}
\end{equation}
because it discards the momentum information from the coherent-state 
measurement. In other words, we consider only ``pure'' measurements. 
The purity of measurement outcome $\alpha$ is defined in terms of the POVM element $\Pi_\alpha=K_\alpha^\dagger K_\alpha$
as \cite{van2017photodetector}
\be
{\rm Purity}_\alpha=\frac{\Tr (\Pi_\alpha^2)}{\left[\Tr(\Pi_\alpha)\right]^2},
\ee
and so we require ${\rm Purity}_\alpha=1$ for all $\alpha$.

\subsection{Coherent-state measurements}

Constructing pure states with the property that both $\expect{X^2}$ and $\expect{P^2}$ are bounded is, fortunately, straightforward on $L^2(\RR,\dd x)$, as the ubiquitous standard harmonic oscillator eigenfunctions have just that property.

Any linear combination
\be
h_{a,b}:= aX_1^2 +b P_1^2
\ee
with $a,b$ real and positive is equivalent to the Hamiltonian of some harmonic oscillator [which is not necessarily the actual Hamiltonian of our system!]
and so has eigenvalues $\lambda_n$ and eigenstates $\ket{\psi_n}$ for $n\in\NN$ in the  Hilbert space $H_1=L^2(\RR,\dd x)$.
Namely, we just define an effective mass $m$ and an effective frequency $\omega$ by
\bea
m&=&\frac{1}{\sqrt{2b}},\nonumber\\
\omega&=&2 \sqrt{ab},
\eea
which then determine the familiar eigenvalues as
\be
\lambda_n=(n+1/2) \omega = 2(n+1/2) \sqrt{ab}
\ee
with the eigenstate $\ket{\psi_0}$ determined by
\be
[\sqrt{a} X_1 +i \sqrt{b} P_1]\ket{\psi_0}=0.
\ee
Given properly normalized eigenstates $\{\ket{\psi_n}\}$ we can then define a coherent state $\ket{\alpha}$ for any complex number $\alpha$ as
\be\label{cohdef}
\ket{\alpha}=\exp(-|\alpha|^2/2)\sum_n \frac{\alpha^n}{\sqrt{n!}} \ket{\psi_n}.
\ee
The projectors onto these coherent states define a POVM 
\be
\frac{1}{\pi}\int_{\CC} \proj{\alpha}\, \dd^2 \alpha=\openone_{H_1}.
\ee
This POVM defines a simultaneous measurement of $X_1$ and $P_1$ with resolutions on the order of
\be
L_0=\frac{1}{\sqrt{m\omega}}=\sqrt{\frac{\sqrt{b}}{\sqrt{a}}}
\ee
for $X_1$ and on the order of $1/L_0$ for $P_1$.

The above construction applies directly to Case 1. Given that in Case 2 we define a Neumark extension that takes us from $L^2(\Omega_2,\dd x)$ to $L^2(\Omega_1,\dd x)=H_1$, the same construction works for Case 2 as well.

For Case 3, the same Neumark extension works once again, but the self-adjoint extension of the momentum operator gives a more interesting result. Namely, once we have defined the orthonormal basis $\{\ket{p_m}\}$ of momentum eigenstates (\ref{pm}) we can use the projector $\openone_3:= \sum_m \proj{p_m}$
to convert the POVM $\{\frac{1}{\pi} \proj{\alpha}\}$ on $L^2(\Omega_1,\dd x)$
to a POVM on $L^2(\Omega_3,\dd x)$ by defining
\be
\Pi_\alpha :=  \openone_3  \frac{1}{\pi} \proj{\alpha}  \openone_3,
\ee
which satisfies
\be
\int_{\CC} \Pi_{\alpha}\,\dd^2\alpha = \openone_3 \openone_{H_1} \openone_3=\openone_3,
\ee
simply because $\Omega_3 \subset \Omega_1$. Thus, the POVM $\{\Pi_{\alpha}\}$ describes a physically acceptable simultaneous measurement of $X_3$ and $\tilde{P}_3$.

Since Case 4 is isomorphic to the self-adjoint extension of Case 3 with position and momentum interchanged, and since coherent states are mapped to coherent states by interchanging the roles of $X$ and $P$, the same construction we just described works for Case 4 as well.

Since the Neumark extension we use for Case 5 takes us to Case 4, the same construction works there as well.

Thus, in the first 5 cases we can define a coherent-state measurement
that describes a simultaneous measurement of both coordinate and momentum, such that expectation values of $X^m$ and $P^m$ are bounded. (Case 6 presents no problems in that regard, as the operators $X_6$ and $P_6$ are bounded and their eigenstates are physical.)
 
Certain {\em discrete} countable sets of coherent states are known to have interesting properties \cite{daubechies1986painless} that may be worth mentioning in our context. First consider a lattice on the complex plane consisting of points $\alpha_{m,n}:=ma+i nb$ for some fixed real $a,b>0$ and $m,n\in \ZZ$.
If we take the set of coherent states $S_{a,b}:=\{\ket{\alpha_{m,n}}\}$, viewed as elements of $H_1=L^2(\RR,\dd x)$ then it is known that this set is not complete for $ab>2\pi$. The border case, $ab=2\pi$ is called the von Neumann lattice, and the set $S_{a,b}$ is complete (i.e., they span $H_1$). This remains true if we remove exactly one element (any one) of the set, but not when we remove two.
Given the set $S_{a,b}$ we are interested in the operator
\be
\Sigma_{a,b}:=\sum_{m,n} \proj{\alpha_{m,n}}.
\ee
If this operator were a nonzero multiple of the identity operator on $H_1$, then we would have a POVM (by simply renormalizing).
However, for the von Neumann lattice one can show that the spectrum of the operator $\Sigma_{a,b}$ contains the number zero, and so it can't be a nonzero multiple of the identity. We need a more dense set of lattice points (i.e., have $ab<2\pi$) to ensure the spectrum to remain strictly positive. Ref.~\cite{daubechies1986painless}
considered the special case of $ab=2\pi/N$ for $N>1$ integer. It shows that
$\Sigma_{a,b}$ then yields a {\em frame}, i.e., there are two positive constants $A$ and $B$ such that
\be
A\openone_{H_1} \leq \Sigma_{a,b}\leq B \openone_{H_1},
\ee
where we pick $A$ as large as possible and $B$ as small as possible.
It is clear that the ratio $B/A$ is a decreasing function of $N$, with $B/A\rightarrow 1$ for $N\rightarrow \infty$. Thus, in that limit we find a POVM, as we should be approaching the coherent-state measurement (for which $A=B=\pi$).

 \section{Conclusions}

 We discussed various mathematical aspects of Hilbert spaces that 
 are not often treated in quantum physics textbooks but do have physical consequences. We focused on the most basic pair of observables, position and momentum, or, to be more precise, generalized coordinates and the associated momenta. We classified the operators one can define for these physical quantities and found 7 basic pairs. We also showed that knowing these operators by no means determines how corresponding measurements are to be described. In fact, realistic measurements are never  measurements of just position or of just momentum, but some combination of both.
 Coherent-state measurements form an important example of this type of measurement.
 
 \section*{Appendix}
Here we collect a number of relevant subjects, spread across the
physics and mathematics literatures, with illustrative examples, which we try to make as simple as possible and
which are relevant to this manuscript.
 
 \subsection{$L^2(\RR,\dd x)$ is a separable Hilbert space}
 Let us see why the standard Hilbert space of square-integrable functions on the real axis (the $x$-axis), $L^2(\RR,\dd x)$ is separable. One very convenient basis is furnished by the eigenfunctions of the harmonic oscillator Hamiltonian (centered around $x=0$) with some arbitrary positive mass $m$ and some arbitrary positive frequency $\omega$. The eigenfunctions with energy $(n+1/2)\hbar 
 \omega$ for $n\in \NN$ form, as is well known, a complete orthonormal basis. Note the Hilbert space $L^2(\RR,\dd x)$ can be used to describe any quantum particle moving in 1D, and the (arbitrary) choice of basis  implies neither that the actual Hamiltonian is that of the harmonic oscillator, nor that the particle has mass $m$. 
 
 The set of vectors in $H$ consisting of all vectors that are superpositions of a {\em finite} number of eigenfunctions with {\em rational} coefficients is a countable dense set.
This fact is what makes   $L^2(\RR,\dd x)$ separable.

\subsection{Simple example of Neumark extension}
Here we reverse-engineer the Neumark extension and show there are two physical interpretations of this procedure in the simplest nontrivial case.
This example is a simpler version of the extension discussed in 
Section~\ref{section:case2}.

Consider the 4-dimensional Hilbert space $H'=\CC^4$ and an orthonormal basis $\{\ket{n}\}_{n=1\ldots 4}$.
This basis gives us a set of four orthogonal projectors
$\{\proj{n}\}_{n=1\ldots 4}$ that sum to the identity: $\sum_n \proj{n}=\openone_{H'}$. 
Now consider a 2-dimensional subspace $H$ such that for all $n$
\be\label{Pin}
\Pi_n:=\openone_{H} \proj{n} \openone_{H}\neq 0.
\ee
Thus, $\{\Pi_n\}_{n=1\ldots 4}$ is a set of four nonzero operators on $H$, obtained by restricting the four projection operators $\proj{n}$ to the smaller Hilbert subspace $H$. 
It is easy to see that the operators $\Pi_n$ are all positive operators and that they sum to the identity on $H$,
\be\label{N4}
\sum_{n=1}^4 \Pi_n =\openone_{H},
\ee
which follows straight from (\ref{Pin}) and the fact that $H \subset H'$. These four operators thus form a POVM on $H$ and thus describe a generalized measurement on the quantum system described by $H$.

The Neumark extension theorem states that one can always run this procedure in reverse, starting with the four positive operators $\Pi_n$ on $H$ that satisfy (\ref{N4}) and ending up with four orthogonal projectors $\proj{n}$ onto $H'$.

The relation between $H=\CC^2$ and $H'=\CC^4$ can be understood in two physically distinct ways.
If we view $H$ as a subspace of $H'$ (which is what we have assumed so far) then we can view our physical system as being one system described by a 4-dimensional Hilbert space, but where we restrict attention to initial superpositions of just the two basis states that span $H$.
But we can also view $H'$ as being given as $H'=H\otimes \CC^2$.
Then we have two different physical systems, each described by a 2-dimensional Hilbert space: our system is a qubit, and the additional system is also a qubit, called an ancilla qubit. Here there is no restriction on the initial state of our system, but there is one on the initial state of the ancilla qubit (typically, we take the ancilla to start in a fixed state $\ket{0}$).

\subsection{Hermitian vs.\ self-adjoint}

We briefly discuss the distinction between a hermitian
operator and a self-adjoint one, because the difference is
not always known to physicists.
(More details can be found in  Refs.~\cite{gieres2000mathematical,bonneau2001self,garbaczewski2004impenetrable}.)
For operators on finite-dimensional Hilbert spaces these are in
fact the same, as they can be applied to any vector in the Hilbert space.
However, a distinction appears for an {\em unbounded} operator $Q$
on an infinite Hilbert space $H$.  In particular, one must then worry about the 
domain of $Q$, which is the set of vectors $\ket{\psi}$ in $H$
such that $Q\ket{\psi}$ is also in $H$. In the case of an unbounded
operator, its domain is always a proper subset of $H$.

Then $Q^\dagger$ is the hermitian conjugate of $Q$ if 
$\smash{\braket{Q^\dagger\phi}{\psi}=\braket{\phi}{Q\psi}}$ \footnote{We use abbreviated notation here: $\langle Q^\dagger\phi|\psi\rangle$ stands for the inner product $(Q^*(\phi), \psi)$ in math notation.}
for every $\ket{\phi}$ and $\ket{\psi}$ in the domain of $Q$.
The hermitian conjugate $Q^\dagger$ has its own domain,
being the set of all $\ket{\phi}$ in $H$, where
$Q^\dagger\ket{\phi}\in H$ 
and $\smash{\braket{Q^\dagger\phi}{\psi}=\braket{\phi}{Q\psi}}$ for all
$\ket{\psi}$ in the domain of $Q$.  The domain of $Q$ is in general a subset of the domain of $Q^\dagger$ \cite{gieres2000mathematical}; it must also be dense in $H$ for $Q^\dagger$ to be well defined.
The operator $Q$ is hermitian if $Q^\dagger=Q$,
but it is self-adjoint under the additional condition that
the domains of $Q$ and $Q^\dagger$ are equal.
For an operator to be an observable in quantum mechanics,
it must be self-adjoint, not merely hermitian. (For example, only then is its spectrum guaranteed to be real.)

The momentum operator $P_3=-i \dd/\dd x$ on the finite interval in the standard textbook case of a particle in a box described by wave functions $\psi(x)$ with $\psi(0)=\psi(L)=0$ is an example
of an operator that is Hermitian but not self-adjoint.

To see this, note 
the condition for being Hermitian is that
\bea
  \label{selfadjextpre}
  0=\bra{\phi}P_3^\dagger\ket{\psi} -\bra{\phi}P_3\ket{\psi}
  &=&
  i\hbar\int_{0}^{L} 
  \Big[\psi \phi'^*
  +\phi^*\psi'\Big]\,\dd x
  \\&=&
  i\hbar
  \Big[\phi^*(L)\psi(L)-\phi^*(0)\psi(0)\Big],\nonumber
\eea
and this quantity indeed vanishes because the function $\psi$ here vanishes on
the boundaries. But we also see there is no restriction necessary on the functions $\smash{\phi(x)}$ in the domain of $\smash{P_3^\dagger}$. Thus, the domain of $\smash{P_3^\dagger}$ is strictly larger than that of $\smash{P_3}$, which means the latter operator is not self-adjoint.

However, Dirichlet boundary conditions are too restrictive
in (\ref{selfadjextpre}).  If we only require
\be\label{bcD}
  \psi(L)=e^{i\theta_0}\psi(0), \quad
  \phi(L)=e^{i\theta_0}\phi(0),
\ee
then Eq.~(\ref{selfadjextpre}) still vanishes.
The domain of $P_3$ has grown, and the domain of $P_3^\dagger$
has shrunk, and their domains end up being the same.  This is an example of a \textit{self-adjoint extension}
of $P_3$ to a self-adjoint version $\tilde{P}_3$.
Eq.~(\ref{bcD}) is a whole family of possible boundary conditions, including periodic
($\theta_0=0$) and antiperiodic ($\theta_0=\pi$) conditions.

\subsection{The finite interval}
In the main text we briefly mentioned there are three different physical situations described by the two possible extensions of the momentum operator $-i \dd/\dd x$ on the finite interval $[0,L]$. The Neumark extension pertains to a particle inside the infinite square well (as an approximation to two high but finite barriers), while the family of self-adjoint extensions pertains to a charged particle moving on a circle with either a zero or a nonzero magnetic flux. These three cases differ mathematically and physically, and so one cannot simply transfer conclusions from one case to the other. Some confusions about a ``particle in a box'' might arise from an illegitimate transfer of intuition.

We start with an example of a very simple wave function that illustrates the different treatments needed in the three different physical situations. The stark physical differences are also reflected in the mathematical question of the proper domain of the momentum operator. 
\subsubsection*{The constant wave function}
Consider this wave function
\bea
\psi(x)&=&\frac{1}{\sqrt{L}}  \,\,\, {\rm for} \,\,x\in [0,L]\nonumber \\
\psi(x)&=&0   \,\,\, {\rm for} \,\, x\notin [0,L]
\eea
Clearly, this wave function is an element of $L^2([0,L],\dd x)$.
This does not quite mean yet that it describes a physical state.

First consider the Neumark extension to  $L^2(\RR,\dd x)$. 
When viewed as a wave function in $L^2(\RR,\dd x)$
$\psi$ is not in the domain of $P_1=-i \dd/\dd x$ since, obviously, the derivative is not defined at $x=0$ and $x=L$. We can evaluate the Fourier transform of $\psi$, which is proportional to the sinc function, $\tilde{\psi}(k)\propto \sinc(kL)$. If we use that function to evaluate expectation values of $P_1^n$ for positive integers $n$, we see that even for $n=1$ a slight problem is encountered, in that the integral 
\be
I_n=\int_{\RR}
 k^n \sinc(kL)\, \dd k \ee
for $n=1$ does not converge absolutely. Any regularization procedure that is symmetric under $k\mapsto -k$ does yield the physically reasonable result $I_1=0$. But for $n=2$ we find a divergent integral $I_2$, so that $\langle P_1^{\,2}\rangle$ diverges.

For the particle on a circle and using the self-adjoint extension with $\theta_0=0$ [that is, using the boundary condition $f(L)=f(0)$
to define the domain of $\tilde{P}_3$] we see that the constant wave function $\psi$ is in the domain of the momentum operator. Indeed, it is the eigenstate of zero momentum. Thus, all moments $\langle\tilde{P}_3^{\,n}\rangle$ vanish, a result very different from what we found for $P_1$!

When we use a nonzero value of $\theta_0$ the constant wave function is not in the domain of $\tilde{P}_3$. We can, however, expand $\psi$ in terms of the eigenstates $\ket{p_m}$ of $\tilde{P}_3$, as defined in Eqs.~(\ref{pmv}) and (\ref{pm}). 
We find
\be
\psi(x)=i\sum_m  \frac{\exp(i\theta_0)-1}{(\theta_0+2\pi m)} \frac{\exp(ik_m x)}{\sqrt{L}},
\ee
where the right-hand side now is in the Hilbert space by construction.
(The convergence is not point-wise, but in the $L^2$ norm.) 
The expectation value $\langle\tilde{P}_3^n\rangle$
is formally calculated by using the right-hand side expansion, which yields
\be
\expect{\tilde{P}_3^n}=4 \frac{\sin^2(\theta_0/2)}{L^2}\sum_m
(\theta_0+2\pi m)^{n-2}.
\ee
The sum diverges for $n\geq 2$, and for $n=1$ converges only conditionally, not absolutely.

This case of the constant wave function can be used as an additional ``paradox'' of the type discussed in \cite{gieres2000mathematical}, in the sense that a mathematical property that at first sight might seem unimportant and ``merely a technical detail'' does have physical consequences.

As a concluding side remark, we note that the limit of $\theta_0\rightarrow 0$ is well-defined when considering the eigenfunction of $\tilde{P}_3$ with $m=0$, but behaves badly if we would use, by mistake, the constant wave function $\psi$.
\subsubsection*{Particle in a box}
Let us first consider a ``particle in a box'' as an idealized limit of a particle moving on the entire $x$ axis in the presence of a ``nice'' potential $V(x)$ that has two ``large'' local maxima around $x=0$ and $x=L$ but is ``small'' in between.
From the eigenvalue equation for the corresponding Hamiltonian we see that
an eigenstate $\psi_n(x)$ is $N+2$ times (continuously) differentiable but not more if $V(x)$ is $N$ times (continuously) differentiable but not more. This implies that the operator $P_1^{N+2}$ is well defined for such a $V$ but $P_1^{N+3}$ is not. The idealized limit, annoyingly, does not even belong to the $N=0$ case, because $V(x)$ itself is not well defined (finite) for $x\leq 0$ and $x\geq L$.

The standard textbook solution for a particle in a box does manage to make sense of eigenstates of the Hamiltonian by imposing zero Dirichlet boundary conditions, but one should realize that the self-adjoint extension of the operator $\tilde{P}^2:=-\dd ^2/\dd x^2$ thus obtained is, in spite of suggestive notation, {\em not} the square of a self-adjoint momentum operator $-i \dd/\dd x$ \cite{garbaczewski2004impenetrable}: the different boundary conditions needed to make these two operators self-adjoint are not compatible (see the discussion in the section on Hermitian vs.\ self-adjoint operators).  

The problem with the seemingly attractive idea of using the square of the self-adjoint extension $\tilde{P}_3$ in the Hamiltonian is that the boundary conditions then prevent the position operator to be defined on the same domain. In contrast, the boundary conditions used for the textbook solution (wave functions vanishing at $x=0$ and $x=L$) do allow the position operator to be defined.  For a different perspective, constructing a self-adjoint Hamiltonian from a non-self-adjoint momentum operator, see \cite{kim2024}.
(Of course, all these problems disappear when using a well-behaved $V(x)$.)

In order to discuss the next two ``paradoxes'' we find it much more convenient to place the boundaries at $x=\pm L/2$.
The question of the proper domain of operators, which we briefly illustrated above for the constant wave function, also plays a role in  Example 7 of Ref.~\cite{gieres2000mathematical}, which we discuss now.
Suppose we wish to calculate the variance of the (kinetic) energy in a given state $\ket{\psi}$ with wave function proportional to $L^2-4x^2$ for a particle in a box on $[-L/2,L/2]$. It may seem we need to calculate $\smash{\expect{\tilde{P}^4}}$ (and then, seemingly, we would get the result 0). 
However, one condition for $\tilde{P}^4$ to be self-adjoint is
\bea
  &&\bra{\phi}(\tilde{P}^\dagger)^4\ket{\psi} -\bra{\phi}\tilde{P}^4\ket{\psi}
  \\&&\hspace{5mm}= 
  \int_{-L/2}^{L/2}
  \Big[\psi{\phi''''}^*
  -\phi^*\psi''''\Big]\, \dd x
  \\&&\hspace{5mm}= 
  -
  \Big[\psi{\phi'''}^*-\psi'{\phi''}^*-\phi^*\psi'''+{\phi'}^*\psi''\Big]_{-L/2}^{L/2}
  \\&&\hspace{5mm}=  0.
\eea
That is, the domain of $\tilde{P}^4$
is functions in the Hilbert space $H$, such that their fourth derivative
is also in the domain, and the wave function and the second
derivative vanish at the boundaries.
However, the wave function for our state $\ket{\psi}$ has a constant (nonzero) second
derivative, so it is not in the domain of $\tilde{P}^4$.

On the other hand, the operator $\tilde{P}^2$ is well defined and self-adjoint, and we can thus calculate the inner product of the vector $\tilde{P}^2 \ket{\psi}$ with itself. This procedure does yield a sensible (and correct) nonzero value for the variance of energy in that state.

Here is another ``paradox'' that illustrates the same point.
A particle of unit mass occupies an infinite square well, which occupies
$[-{L}/{2},{L}/{2}]$.
Then calculate $\langle[X_3^{\,4},\tilde{P}^4]\rangle$ with respect to any energy eigenstate.
One person says that $\tilde{P}^4$ is proportional to the square of energy, which is well
defined.  Thus, with respect to the $m$th eigenstate,
\be
  \expect{X_3^{\,4}\tilde{P}^{4}-\tilde{P}^{4}X_3^{\,4}}
  =p_m^{\,4}\expect{X_3^{\,4}-X_3^{\,4}}  =0.
\ee
However, another person interprets the momentum operator as a derivative
acting on the right, and the position operator as multiplication by $x$,
acting to the right, and obtains
\be
  \expect{[X_3^{\,4},\tilde{P}^4]}
  =4\pi^2.
\ee
Who is right?

The first answer is correct, and there are different ways to show this.
A mathematical physicist would say, again, 
that the domain of $\tilde{P}^4$ (as a self-adjoint operator)
contains functions whose second derivatives vanish at the boundaries.
So while any energy eigenfunction $\psi_m(x)$ is in the domain of $\tilde{P}^4$,
the functions $x^4\psi_n(x)$ are not in the domain of $\tilde{P}^4$, because
$(x^4\psi_m)''\neq 0$ at $x=\pm{L}/{2}$.
Thus, it is not allowed to let $\tilde{P}^4X_3^{\,4}$ act to the right on these functions.
Instead, this combination should, for example, act to the left, and the
expectation value vanishes.

A more practical physicist would regularize the infinite square well 
into its finite counterpart (replacing $X_3$ with $X_1$), 
and then let the well depth go to infinity.
Doing the calculation this way, one finds that the region inside
the well contributes $4\pi^2$ to the expectation value, as before.
But the classically forbidden regions outside the well also contribute
$4\pi^2$. It is the step discontinuities themselves that give a contribution
of $-8\pi^2$, for an overall vanishing value of the expectation value.
In this point of view, it is not the illegal operation of acting to the right
that is the problem, but that contributions from outside the well were missed.

\subsection{$\RR^3$}
The Hilbert space for a particle moving in 3D space is $L^2(\RR^3,\dd \vec{r})$.
 If we view this Hilbert space as a tensor product $L^2(\RR,\dd x)\otimes L^2(\RR, \dd y)\otimes L^2(\RR, \dd z)$ then we will encounter no new problems with the definitions of momentum operators in the $x,y,z$ directions, as each case reduces to the standard Case 1. If we use spherical coordinates $r,\theta,\phi$ [as we certainly should sometimes!] then we may seem to run into some trouble with radial momentum, the momentum canonically conjugate to the variable $r\in [0,\infty)$. This is an example of Case 2, for which no self-adjoint extension exists and where the Neumark extension to the whole real axis (by allowing for negative values of $r$) would make no physical sense.  

 Of course, there is no real problem here. The spherical coordinate system is itself not flawless, since this single set of coordinates cannot cover all of $\RR^3$. 
 On the $z$ axis the azimuthal angle $\phi$ [using physicists' conventions]
 is not defined and in the origin $r=0$, neither $\theta$ nor $\phi$ are defined. Normally, when we use spherical coordinates, for example to describe the motion of Earth around the Sun, we make sure that the physical effects we wish to describe take place outside those coordinate singularities. For example, Earth is always taken to move in the $x,y$ plane and the Sun occupies the origin, so Earth will never occupy the $z$ axis.

 If we discuss the quantum mechanics of, say, the hydrogen atom in spherical coordinates, we similarly assume the proton occupies the origin. More importantly (to cover cases where no particle occupies the origin), the only wave function nonzero at the origin has zero angular momentum and hence has no dependence on either $\theta$ or $\phi$, thus bypassing the problem of these angles not being defined. 
 
For points on the $z$ axis, where $\phi$ is not defined, we note that a $\phi$-dependent wave function occurs only if $m\neq 0$, and for those wave functions there is always at least one factor of $\sin\theta$, thus making the wave function vanish on the $z$ axis. (Physically speaking, one may associate a vortex with $m\neq 0$). Of course, the volume element in spherical coordinates contains a factor of $\sin\theta$ as well, but since the gradient operator contains terms with a $1/\sin\theta$ factor, this alone would not suffice to get rid of problems on the $z$ axis.

\subsection{Coherent-state measurements in optics}
A standard measurement in optics (classical or quantum) consists of combining on a 50/50 beam splitter an input signal with a so-called local oscillator, a beam of light with known frequency and average intensity (or photon number per unit of time). One then measures the difference in the intensities at the two output ports of the beam splitter over some finite amount of time $T$.
If the frequencies of the signal mode and local oscillator mode coincide then this is called balanced homodyne detection, if the frequencies differ (by much more than $1/T$), heterodyne detection. 
It provides an example of a POVM measurement on {\em one} signal input mode implemented by a direct measurement on {\em two} output modes. The local oscillator field thus serves as an ancilla (described by an infinite-dimensional Hilbert space). This then gives a concrete and straightforward example of enlarging the Hilbert space and obtaining a Neumark extension.
Refs.~\cite{goetsch1994linear,wiseman1996quantum} show how the heterodyne measurement yields the coherent-state measurement in the limit of large $T$.

Here we briefly outline part of Ref.~\cite{jackson2023sim1}, which uses a different approach and describes a simultaneous continuous measurement of $X$ and $P$, which then also reduces to the coherent-state measurement in the limit of large measurement time $T$. 
Starting from an interaction Hamiltonian between a system and a meter one obtains an (unnormalized) Kraus operator (where we use a dimensionless time $t$ and dimensionless self-adjoint operators $X$ and $P$ for our system) 
\be
K_{dt}=\exp\Bigl(X \dd W_X+ P \dd W_P -(X^2+P^2) \dd t\Bigl).
\ee
Here $\dd W_k$ for $k=X,P$ are two independent Wiener increments with  normal distribution with average zero and $(\dd W_k)^2=\dd t$.  
This Kraus operator describes a simultaneous (weak) measurement of $X$ and $P$ during a very short time interval $\dd t$. The EXP map maps the Lie algebra generated by $\{X,P,X^2+P^2\}$ to a Lie group. 
Taking commutators of these 3 elements generates further basis elements of the Lie algebra (viewed as a {\em real} vector space). For example, the commutators of $X^2+P^2$ with $X$ and $P$ yield new basis elements $iP$ and $iX$, respectively ($\pm$ signs being real don't matter). The commutator of $X$ and $P$ gives $i\openone$ and the commutator of $X$ and $iP$ gives $\openone$.
No more elements are generated.
Thus we get a 7-dimensional Lie algebra with
7 basis elements  $\{\openone, i\openone, X, P, iX, iP, X^2+P^2\}$. 

Via the EXP map this Lie algebra then generates a 7-dimensional Lie group, which contains the Kraus operators. Note this is not a unitary group, the usual type of Lie group encountered in quantum mechanics.
The random (weak) measurement outcomes produce a random walk through this  Lie group, starting at the identity operator. In the limit of $T\rightarrow \infty$ the path will always reach a 2-dimensional boundary of that manifold, which contains all projectors onto coherent states (acting on the Hilbert space on which the original operators $X$ and $P$ act). In fact, that boundary is reached exponentially fast. Thus the 
accumulation of weak measurements produces a strong measurement---the heterodyne measurement---in the limit $T\rightarrow \infty$. 
For proofs of these (far from obvious!) statements and further examples (specifically, measurements of non-commuting components of angular momentum), see Refs.~\cite{jackson2023perform,jackson2023sim1,jackson2023sim2}.

 \subsection{Time and phase}
 For the energy eigenfunctions $\ket{n}$ of some specific harmonic oscillator we may define (in physicists' notation)
 \be
 \ket{t}=\sum_{n\in \NN} \exp(-i n \omega t)
 \ket{n}.
 \ee
 The measure $\dd\mu_t:=\proj{t} \dd t$ is a POVM but not a projection-valued measure.
 It thus describes a generalized measurement of what would have to be time, the variable conjugate to energy. We can define a self-adjoint time operator on a larger Hilbert space, exactly as in Case 5, by doubling the Hilbert space by adding an ancilla qubit.
(Of course, trying to define ``time'' as conjugate to an arbitrary Hamiltonian does not make much sense, just as it does not make much sense to try using an arbitrary system, with arbitrary time evolution, as a clock. Here we followed an old tradition and used a pendulum as a clock.)
 
 Defining a phase operator for the free EM field can be phrased in exactly the same way as the above, with the solution again as given for Case 5. One thus can define a POVM describing phase measurement. This has been known since at least 1991 \cite{hall1991}. One can define a self-adjoint phase operator on a doubled Hilbert space, which then has a natural interpretation as a phase-difference operator.

\subsection{Continuum fields in quantum optics}
Here we describe a relatively simple example of how an initial inseparable Hilbert space for a 1D field theory is restricted to a much smaller physical subspace---a Fock space---that is separable.
The title of this subsection is taken from Ref.~\cite{blow1990continuum} where the following procedure is described in more detail and which gives various practical applications of the formalism.

If we quantize the free 1D EM field we may define creation and annihilation operators for each wave vector $k$. One would call $c|k|$ the frequency of the corresponding photon. We define $\omega=ck$ to connect to more standard notation,  but note that our $\omega$ can be negative as well as positive.
In physicsts' notation we then define a delta-function normalized ket
\be
\ket{\omega}:=a^\dagger(\omega)\vac,
\ee
with $\vac$ denoting the vacuum state, containing zero photons.
This, of course, does not describe a physical state. Instead, a physical (pure) single-photon state is of the form
\be
\ket{\Phi}:=\int_\RR \Phi(\omega) a^\dagger(\omega)\,\dd \omega \vac,
\ee
with normalization condition
\be
\int_\RR |\Phi(\omega)|^2\,\dd \omega =1.
\ee
The normalization condition shows that the mode functions $\Phi$
together form a Hilbert space $L^2(\RR, \dd \omega)$, which is separable. We may thus define an orthonormal basis set of functions $\Phi_n(\omega)$
for $n\in\NN$ for that Hilbert space such that
\be
\int_\RR  \Phi^*_n(\omega) \Phi_{n'}(\omega)\,\dd \omega =\delta_{nn'}.
\ee
(The point of Ref.~\cite{blow1990continuum} is that in this way we have discretized the modes without having defined a finite box whose boundary conditions would lead to a discrete set of allowed frequencies.)
For each $n$ we have now an infinite-dimensional Hilbert space spanned by the Fock states $\ket{m}_n$, where
\be
\ket{m}_n=\frac{\left(a_n^\dagger\right)^m}{\sqrt{m!}} \vac,
\ee
with
\be
a_n^\dagger:= \int_\RR  \Phi_n(\omega) a^\dagger(\omega)\, \dd \omega .
\ee
The space $L^2(\RR,\dd\omega)$ of mode functions thus corresponds to the subspace of all single-photon states, i.e., all states with exactly one excitation in one mode $n$ and zero in all others.
Note here $\vac$ is defined as the (unique) state that satisfies
\be
\forall n:\,\, a_n \vac=0.
\ee
Restricting to the subspace consisting of states with bounded total number of photons yields a separable space, called Fock space. (Completing Fock space yields a Hilbert space, separable by definition.)
The tensor product of all countably many Fock spaces is, however, inseparable. We continue this theme in the next section in a more general context.

\subsection{Quantum field theory}

Essentially all Hilbert spaces in quantum mechanics are separable.
Recall that a metric space is separable if it has a countable, dense subset
(and thus a countable, orthonormal basis)---for example, $L^2(\RR^N)$
has products of $N$ harmonic-oscillator eigenfunctions as a basis.
However, quantum field theories typically deal with a space more like
$L^2(\RR^{\NN})$, which comprises infinite sequences of real numbers.
This is not a separable space \cite{fairbairn2004separable}.
Quantum statistics in the thermodynamic limit (e.g., an infinite chain of spins), too, initially yields an inseparable Hilbert space, arising from the infinitely many degrees of freedom one starts with.

Why do we construct a separable Hilbert space, even in those cases where we start with an infinite number of degrees of freedom?
In the context of an inseparable Hilbert space, Talagrand \cite{talagrand2022quantum} gives examples of generalizations of unitary operators,
operating on all dimensions in the same way, that give absurd results.
For example, let $a_k$ be an annihilation operator on the $k$th dimension,
and consider the Bogoliubov-type transformation
$a'_k = a_k\cosh\zeta + \smash{a^\dagger_k}\sinh\zeta$ ($\zeta>0$).
Let $\ket{0}$ and $\ket{0'}$ be the respective ground states of $a_k$ and $a_k'$
($a_k\ket{0}=0$, $a_k'\ket{0'}=0$).
Then the inner product of these ground states involves an infinite product
of quantities which are constant and less than one, so $\braket{0'}{0}=0$.
Witten \cite{witten2021does} puts this argument in general terms, noting that
a change to a finite number of dimensions 
(e.g., by applying a finite number of creation operators) does not alter this argument.
Thus, all of the Fock states in the $\ket{0}$ basis are orthogonal to all
of the Fock states in the $\ket{0'}$ basis.
(For another (much simpler) example consider a unitary phase shift
applied to each local degree of freedom $n$, $U_n=\exp(i\phi_n)\openone_n$ with the phase shift $\phi_n$ defined relative to a local phase reference. Clearly, if $\sum_n \phi_n$ does not converge, then neither does the tensor product $\otimes_n U_n $.)

There is also a famous result known as Haag's theorem
\cite{haag1955,roman1969,streater2000pct,talagrand2022quantum}.
In brief, the theorem says that, under certain (reasonable) assumptions,
if a quantum field theory 
is unitarily equivalent to a free field theory,
then the original field theory must also be free.
This is an apparently absurd result, because interacting quantum field theories exist. 
The proper way to handle these unitarily inequivalent representations is within the algebraic formulation of quantum theory, see the textbook \cite{Landsman2017}, or for a brief summary see Section 4 of Ref.~\cite{Ruetsche2011}, which is the only rigorous quantum theory known to handle infinitely many degrees of freedom. Within that theory such representations can be given a natural physical interpretation. For example, in the thermodynamic limit, different phases necessarily correspond to unitarily inequivalent representations.

The problem in both cases lies in the implicit assumption \footnote{
As Talagrand \cite{talagrand2022quantum} says on page 356: 
``Even though [the Hamiltonians for the two field theories] both make perfect sense, it is the (implicit) assumption that they operate on the same state space that does not.''
} that there is
a single state space, which is closed under unitary transformations.
But really, there are an uncountable infinity of subspaces (say, of $L^2(\RR^{\NN})$), each of 
which is separable and each of which describes a different physical situation. The transformation from one to the other would typically require an infinite amount of energy.  
Each subspace is the completion of the Fock basis;
it includes states of arbitrarily many particles \cite{streater2000pct}.
Within each separable subspace, one may do unitary 
transforms---provided that, as a function of dimension, they rapidly 
become the identity.
A transformation (an unphysical one!) as described in the above examples takes one out of 
the separable space, and into a new one.

Alternately, one can take the point of view that all troubles are a 
consequence of the infinite products involved, and one should 
``regularize'' the field theories to a finite $N$, only taking the
limit to infinity at the end of the calculation 
\cite{witten2021does,preskill2019}.
This is more or less what happens in a computer simulation  when one checks that the end result no longer numerically changes with increasing $N$. The simulation of a quantum field theory on a quantum computer \cite{preskill2019,bauer2023quantum} would be an even more direct example of reliance on a separable Hilbert space.

\bibliography{Hspaces}

\end{document}